# Size-dependence and high temperature stability of radial vortex magnetic textures imprinted by superconductor stray fields.

D. Sanchez-Manzano[1,2], G. Orfila[2], A. Sander[1], L. Marcano[3,4,5], F. Gallego[2], M. A. Mawass[3], F. Grilli[6], A. Arora[3], A. Peralta[2], F.A. Cuellar[2], J.A. Fernandez-Roldan[7], N. Reyren[1], F. Kronast[3], C. Leon[2], A. Rivera-Calzada[2], J.E. Villegas[1], J. Santamaria[2], S. Valencia[3,*].

[1] Unité Mixte de Physique, CNRS, Thales, Université Paris-Saclay, Palaiseau, France.

[2] GFMC. Dept. Física de Materiales. Facultad de Física. Universidad Complutense. 28040 Madrid, Spain.

[3] Helmholtz-Zentrum Berlin, Albert-Einstein Str. 15, 12489 Berlin, Germany.

[4] Dept. of Physics, Faculty of Science, University of Oviedo 33007 Oviedo, Spain.

[5] Center for Cooperative Research in Biomaterials (CIC biomaGUNE), Basque Research and Technology Alliance (BRTA), Paseo de Miramón 194, Donostia-San Sebastián 20014, Spain

[6] Karlsruher Institut für Technologie, Institut für Technische Physik, 76344 Eggenstein-Leopoldshafen, Germany.

[7] Helmholtz-Zentrum Dresden-Rossendorf e.V., Institute of Ion Beam Physics and Materials Research, 01328 Dresden, Germany;

*Corresponding author: sergio.valencia@helmholtz-berlin.de

## Abstract

Swirling spin textures, including topologically non-trivial states, such as skyrmions, chiral domain walls, and magnetic vortices, have garnered significant attention within the scientific community due to their appeal from both fundamental and applied points of view. However, their creation, controlled manipulation, and stability are typically constrained to certain systems with specific crystallographic symmetries, bulk, or interface interactions, and/or a precise stacking sequence of materials. Here, we make use of the stray field of $YBa_2Cu_3O_{7-\delta}$ superconducting microstructures in ferromagnet/superconductor hybrids to imprint magnetic radial vortices in permalloy at temperatures below the superconducting transition temperature ($T_C$), a method easily extended to other ferromagnets with in-plane magnetic anisotropy. We examine the size dependence and temperature stability of the imprinted magnetic configurations. We show that above $T_C$, magnetic domains retain memory of the imprinted spin texture. Micromagnetic modelling coupled with a SC field model reveals that the stabilization mechanism leading to this memory effect is mediated by microstructural defects. Superconducting control of swirling spin textures below and above the superconducting transition temperature holds promising prospects for shaping spintronics based on magnetic textures.

**Introduction**

The rise of spintronics has stimulated the interest in topologically non-trivial spin configurations, such as skyrmions[1–3], merons[4–6], and magnetic (radial) vortices[7–9], promoting the search for new materials, methods, and/or configurations in which these structures can be created, stabilized, and controlled. There have been significant advances in our understanding of the physics governing the formation of these non-trivial magnetic domain configurations. However, a method for creating and stabilizing complex spin textures, and applicable to a large variety of compounds is yet missing. Recently, a new approach based in the use of hybrid superconductor/ferromagnet (SC/FM) microstructures has demonstrated the possibility to generate and control swirling spin textures [10, 11].

Superconductivity and ferromagnetism are two electronic ground states which despite their antagonistic character, may become synergistic in superconductor/ferromagnet hybrids. They yield exciting responses as, for example, the recently demonstrated long range supercurrent and Josephson effects driven by equal spin triplet superconducting correlations[11–13] which coexist peacefully with ferromagnetism.

A wide category of effects in SC/FM hybrids involve the influence of the ferromagnet on the superconducting ground state. This influence is mediated by the stray fields of ferromagnetic domains or magnetic chiral structures (like magnetic vortices, skyrmions or domain walls) on the dissipation properties and critical current characteristics of the superconductor[14–20]. For instance, the presence of a magnetic vortex in the ferromagnetic barrier of a Josephson junction can tailor the supercurrent pathways making it to behave as a SQUID[21] or 0-π SQUID[22].

The opposite, that is the role played by superconducting stray fields in the magnetic ground state of ferromagnets, has comparatively received less attention. To this regard, Palau et al. [23] and Sander et al. [10] have shown the possibility of crafting swirling spin textures in ferromagnetic systems by making use of the magnetic stray fields generated by the trapped flux in structured type II superconductors. Below the superconducting transition temperature, the application and removal of an out-of-plane magnetic field yields the generation of screening supercurrents due to the penetration, pinning and expulsion of magnetic flux quanta[24,25]. Within the mixed state, these supercurrents flow following the geometrical contour of the SC structure [26–28], with a geometry and sense of rotation which depend on magnetic history[10,23]. Supercurrents present after removal of the external magnetic field, give rise to a stray magnetic field whose strength and direction varies locally[29], see Supplementary Information Section 1. In ferromagnetic systems with perpendicular magnetic anisotropy (PMA), the out-of-plane component of the superconductor stray-field can be employed to imprint unusual magnetic textures[10]. The imprint is stabilized by the PMA so that it remains even when supercurrents have vanished for temperatures above $T_C$. On the other hand, Palau et al. showed that in ferromagnetic layers with in-plane the in-plane components of the SC stray-field can be utilized to imprint magnetic domain distributions akin to radial vortices with a lateral size of 20 μm[23]. In these, the in-plane magnetization can point towards or away from the core along radial directions orthogonal to the contour of the superconducting microstructure. This type of magnetization distribution is not energetically favored due to large dipolar energies[30]. Consequently, the disappearance of the SC stray-field above $T_C$ is expected to lead to its relaxation to an energetically more favorable magnetic state, such as a conventional vortex or a multidomain configuration.

Here, we explore how the reduction of the lateral size of the SC structure affects the SC imprint of radial vortex-like magnetic spin textures on $Ni_{80}Fe_{20}$/$YBa_2Cu_3O_{7-\delta}$ (Py/YBCO) hybrids. Finite-difference micromagnetic modelling coupled with the YBCO field modelling indicate that the radially inhomogeneous field distribution of the superconductor enables the imprint of these topologically non-trivial magnetic domain distributions below $T_C$ for lateral sizes down to sub-micrometer. Experimentally we obtain radial vortex-like imprints down to 2 μm most likely limited by the presence of surface defects. Interestingly, although increasing the temperature above $T_C$ leads to the disappearance of the stabilizing SC stray field and the relaxation of the imprinted magnetic domain pattern, the remnant spin texture retains a significant memory of the imprinted state (non-volatile).

The robustness of this state and the origin of this memory effect is discussed in terms of pinning of domain walls by YBCO surface defects, which contribute to stabilize its topology.

**Experiment**

We have fabricated microstructured SC/FM hybrids with square (□) and disc (⊙) shape based on $Fe_{20}Ni_{80}/YBa_2Cu_3O_{7-\delta}$ (Py/YBCO). Two types of SC/FM systems have been investigated. The first system is made of samples for which a continuous Py film has been deposited on top of YBCO structures of different sizes, as in Ref. [23]. The second type consists of samples where the Py has been structured with the same shape but with smaller sizes than that of the SC microstructure underneath (20 μm). From now on we will use the notation (□, ⊙)$SC^{\emptyset}$/$FM^{cont}$ to refer to (square or circular) structures with continuous Py layer and variable SC dot size, and (□, ⊙)$SC^{20}$/$FM^{\emptyset}$ to refer to structures with structured Py on top of 20 μm SC dots, respectively. The superscript Ø indicates the lateral size of the SC or FM in μm as well as the size of imprinted magnetic domains.

A 250 nm thick superconducting YBCO layer was epitaxially deposited on top (001)-oriented Nb-doped $SrTiO_3$ substrates by means of high oxygen pressure (3.4 mbar) d.c. magnetron sputtering at 900°C. Following growth, in-situ annealing in pure oxygen for 30 min at 550°C was performed to ensure an optimal oxygen stoichiometry. Growth conditions, optimized for epitaxial c-axis growth, lead to superconducting films with transition temperature (89 K) close to that of the bulk (92 K). As-grown films are decorated by the presence of CuO surface precipitates characteristic of the high oxygen pressure sputtering growth and difficult to avoid. Precipitates are randomly distributed with diameter sizes varying between ca. 100 and 500 nm, see Figure 1 and Supplementary information of Ref. 10. Superconducting YBCO square and disc structures, with lateral sizes (or diameter) ranging between Ø = 1 μm and Ø = 20 μm, were defined by means of electron beam lithography and etching. A 4nm thick layer of permalloy ($Ni_{80}Fe_{20}$) was deposited by means of magnetron sputtering at room temperature after the structuring of YBCO. All samples were capped with 3 nm of Al to prevent oxidation. A second lithographic process (see methods) followed in those cases where the FM was structured into the same shape as the SC.

Single out-of-plane magnetic field pulses of +100 mT or -100 mT were applied at 50 K (T < $T_C$) to induce the supercurrent distribution that generates a magnetic stray field from the SC. Its impact on the magnetic domain structure of the FM was imaged by means of X-ray Photoemission Electron Microscopy (XPEEM). X-ray Magnetic Circular Dichroism (XMCD), measured at the Fe $L_3$-edge (707 eV), was used as magnetic contrast mechanism (see methods). XMCD images as function of T have been obtained to evaluate the impact of the disappearance of the SC stray field above $T_C$ on the imprinted magnetic domains. All XMCD images were obtained at magnetic remanence i.e., in absence of external magnetic fields.

**Results**

After a zero-field cool process down to 50 K (T < $T_C$) hybrid SC/FM structures with continuous (Figure 1a,b) and structured (Figure 1c,d) Py present a magnetic multidomain state (Figure 1e-h). An out-of-plane magnetic field pulse of -100 mT (see methods) triggers a profound modification of the Py magnetic domain distribution due to the self-field of the SC structure, see Figures 1j-l. The magnetization direction sensitivity of XMCD excludes that the resulting magnetic domain pattern is a magnetic vortex. The XMCD signal distribution for a conventional vortex would look alike to that depicted in Figures 1j-l but under a rotation of ±90 degrees (Supporting Information Section S2). The resulting magnetic domain state, both for square and disc-shaped structures, features a radial vortex-like configuration where the local magnetization is orthogonal to the microstructure contour and points towards its geometrical center[23]. Both, the resolution as well as the in-plane magnetic sensitivity of the experimental set up, prevent "visualization" of the central core predicted by micromagnetic simulations (Supporting Information Section S3). Consequently, from now on we restrict our analysis to the in-plane components of the imprinted magnetization.

In the case of discs, the radial magnetization direction features a continuous and smooth 360° rotation around its center, alike to that reported for radial vortices[9,30]. On the other hand, square structures display magnetic domain walls along the diagonals splitting the magnetic domain state into four equal triangular-shaped head-to-head magnetic domains, where the magnetization direction rotates 90° between adjacent ones. Full reversal of the magnetic domain structure can be achieved by changing the sign of the out-of-plane magnetic field pulse, whereas intermediate states can be obtained by changing its strength[23].

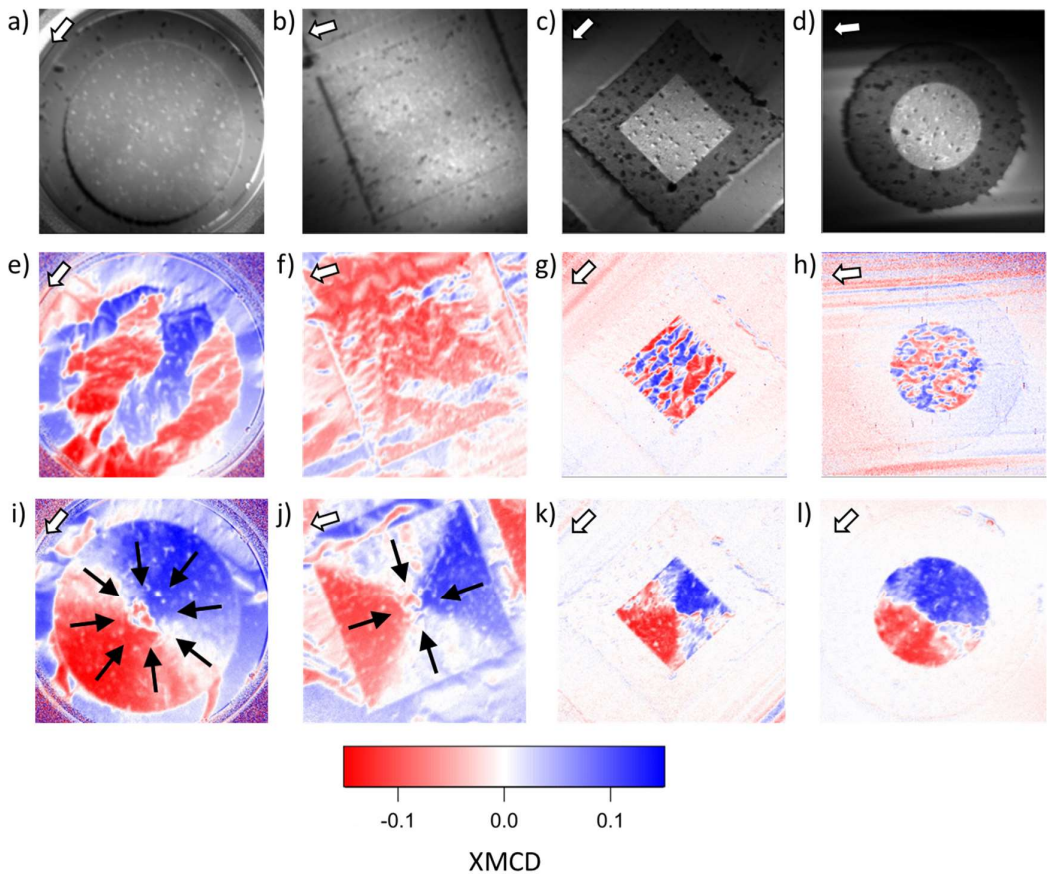

**Figure 1** XAS PEEM images obtained at T = 50 K for hybrid SC/FM structures with continuous Py layer on top a) ⊙$SC^{\varnothing}/FM^{cont}$ and b) □$SC^{\varnothing}/FM^{cont}$ and with structured Py c) □$SC^{20}/FM^{10}$ and d) ⊙$SC^{20}/FM^{10}$. E-h) XMCD images obtained at T = 50 K after a zero-field cool process. I-l) Corresponding XMCD images after an out-of-plane magnetic field pulse of -100 mT. Black arrows indicate the direction of the imprinted magnetization which resembles that of a magnetic radial vortex. White arrows signal the in-coming x-ray beam direction.

The effectiveness of the imprint for (□,⊙)SC$^{\emptyset}$/FM$^{cont}$ and (□,⊙)SC$^{20}$/FM$^{\emptyset}$ structures as function of the imprint size $\emptyset$ is shown in Figures 2 and 3, respectively. XMCD images have been averaged over several similar structures to mask non-magnetic regions linked to surface defects. Both sample systems show a decrease in the efficiency of the imprint of radial vortex magnetization distributions for smaller $\emptyset$. Structures with a continuous Py layer, (□,⊙)SC$^{\emptyset}$/FM$^{cont}$, show a steady decrease of the XMCD strength as the size is reduced. This can be related to the increased weight of non-magnetic regions (XMCD = 0) in the averaging and/or to a non-deterministic imprint. The latter is clear for (□,⊙)SC$^{2.5}$/FM$^{cont}$ structures for which the average XMCD images (Figure 2e,f) show no traces of the SC imprint. For these samples the reduction of the lateral size of the SC dot leads to an overall decrease of the stray field of the superconductor[31,32], see Supplementary Information Figures S1 and S2. Yet, the in-plane fields generated by (□,⊙)SC$^{2.5}$ structures are high enough to align the magnetization of a 4 nm thick Py film.

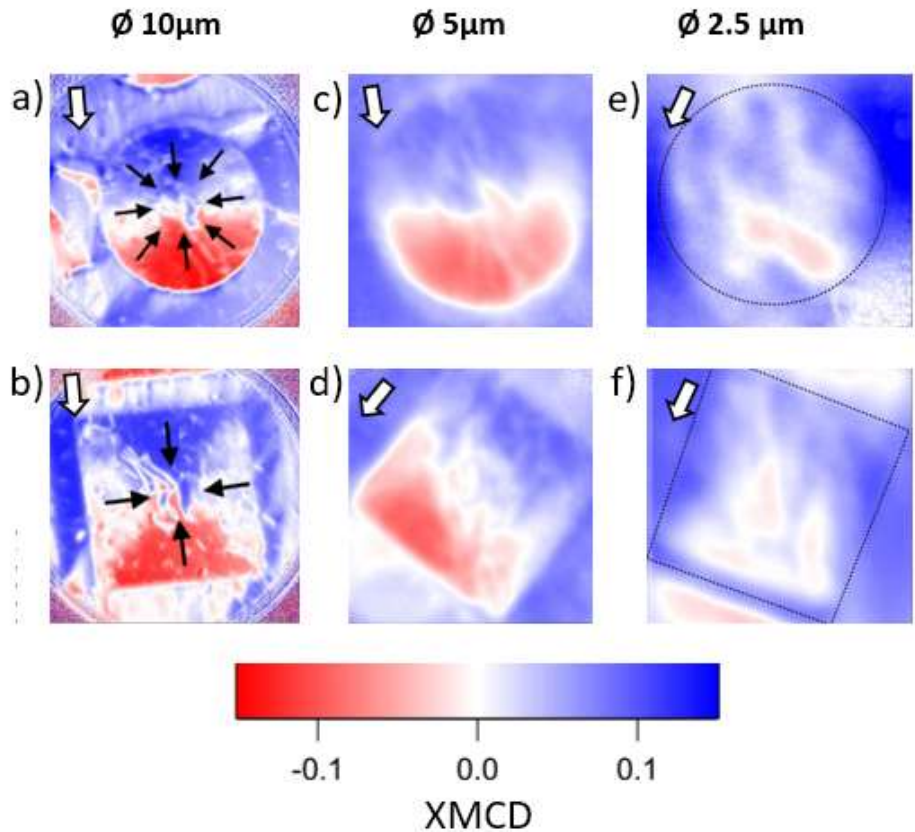

**Figure 2** XMCD images obtained at T = 50 K after a magnetic field pulse of -100 mT for a)-b) (⊙,□)$SC^{10}/FM^{cont}$, c)-d) (⊙,□)$SC^{5}/FM^{cont}$, and e)-f) (⊙,□)$SC^{2.5}/FM^{cont}$. XMCD corresponding to panels c)-f) have been averaged over 10, 7, 8 and 12 similar structures, respectively. Images corresponding to panels a)-b) correspond to a single structure. Black arrows indicate the direction of the imprinted magnetization which resembles that of a magnetic radial vortex domain. White arrows signal the in-coming x-ray beam direction.

The efficiency of the imprint improves when the Py is structured on top of SC dots with the largest stray field (∅ = 20 μm), i.e. for (□,⊙)$SC^{20}/FM^{\emptyset}$ samples, see Figure 3. Such an improvement is evidenced by the fact that XMCD averaged images corresponding to (□,⊙)$SC^{20}/FM^{20}$, (□,⊙)$SC^{20}/FM^{10}$, and (□,⊙)$SC^{20}/FM^{5}$ show a similar XMCD signal distribution, which is expected from a deterministic imprint. The imprint of domains down to ∅ = 5 μm and ∅ = 2 μm is possible for □ and ⊙ structures, respectively, despite the large relative surface area occupied by defects (white areas in Figure 3g).

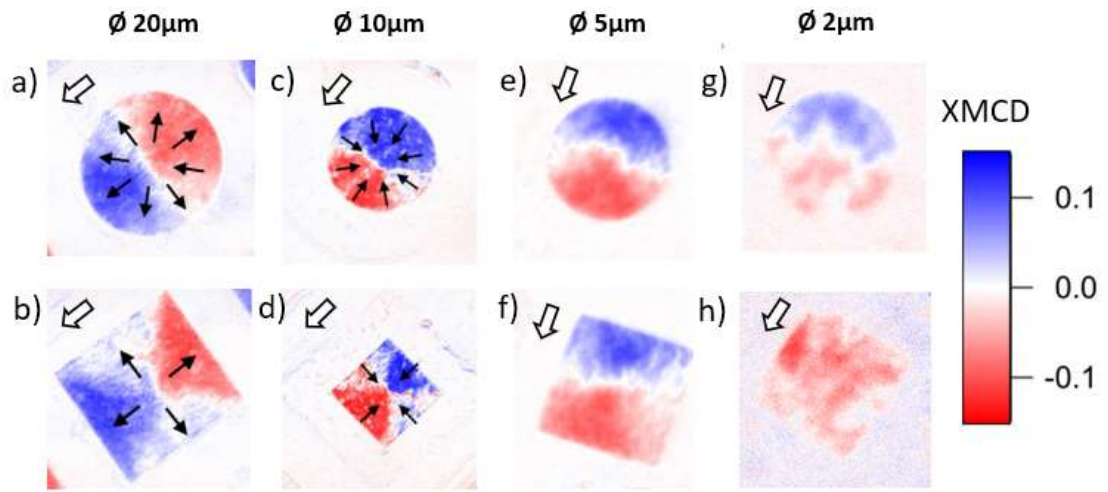

**Figure 3** XMCD images obtained at T = 50 K after a magnetic field pulse of -100 mT for a)-b) (⊙,□)$SC^{20}/FM^{20}$, and after a magnetic pulse of +100 mT for c)-d) (⊙,□)$SC^{20}/FM^{10}$, e-f) (⊙,□)$SC^{20}/FM^{5}$, and g)-h) (⊙,□)$SC^{20}/FM^{2}$ structures. XMCD corresponding to panels a)-b), e)-g) and panel h) have been averaged over 8, 10 and 4 similar structures, respectively. Images corresponding to panels c)-d) correspond to a single structure. Black arrows indicate the direction of the imprinted magnetization which resembles that of a magnetic radial vortex domain. White arrows signal the in-coming x-ray beam direction.

The stability of the imprint as the temperature is increased above the SC transition temperature has been investigated by obtaining XMCD images as function of temperature after a superconducting imprint at 50 K. Figure 4 depicts the data obtained for □$SC^{20}/FM^{cont}$ structures. A total of 10 similar structures have been measured. Similar results have been obtained for □$SC^{20}/FM^{20}$. Inset panels a) to e) depict XMCD images obtained for one of these structures. At 80 K, close to $T_C$ (89 K), there is a partial relaxation of the imprinted magnetic domain state as evidenced by the appearance of dendritic domains and an overall change of the XMCD strength. In between 80 K and 250 K there is little variation. Similar qualitative results have been obtained for all structures measured.

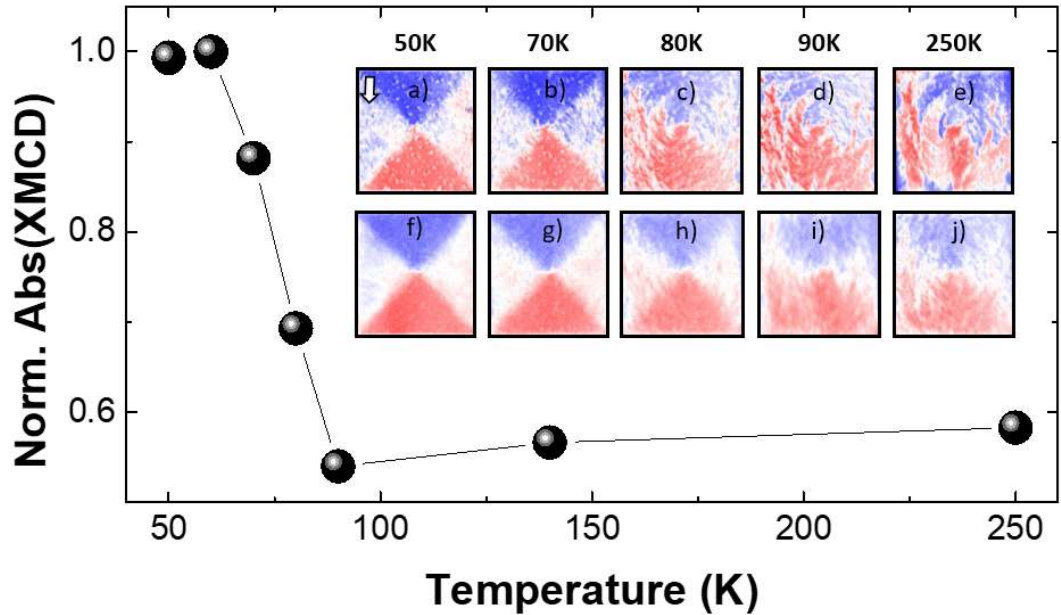

**Figure 4** a-e) and f-j) XMCD images obtained as function of temperature after a magnetic field pulse of -100 mT for ☐$SC^{20}/FM^{cont}$ structures. a)-e) correspond to a single structure. f)-j) are averaged over 10 similar structures. Main panel: Temperature dependence of the absolute XMCD integrated from panels a)-e) over an area defined by the "blue" and "red" triangular domains in a). A white arrow signals the in-coming x-ray beam direction. XMCD scale as in previous figures.

Fine details concerning the relaxation of the imprinted magnetic domain pattern depend on the particularities of each of the 10 systems measured, namely the distribution of defects. To wash out these particularities and reveal their common behavior we depict in inset panels f-j of Figure 4 the corresponding XMCD images averaged over all ten structures. At all temperatures, the average images feature, albeit with some relaxation at higher temperatures, the magnetic domain pattern imprinted at 50 K. Main panel of Figure 4 shows the temperature dependence of the absolute value of the averaged XMCD signal, |XMCD|, integrated over an area defined by blueish (XMCD > 0) and reddish (XMCD < 0) domains at 50 K (Figure 4f). As T increases, the |XMCD| signal decreases, reaching about 50% of its initial value at T = 90 K and remaining roughly constant up to 250 K. This reduction of the XMCD signal is associated with the decrease and final disappearance of the SC stray-field at $T_C$ leading to the relaxation of the imprinted magnetic state. The non-vanishing XMCD signal above $T_C$ indicates that the imprinted domains do not relax to a conventional vortex pattern configuration once the SC stray-field vanishes (Supporting Information Section 3), suggesting the relevance of inhomogeneous microstructure as stabilization mechanism of the imprinted magnetic configuration.

**Discussion**

The imprint of radial magnetic vortex configurations in samples with a continuous Py layer is not observed for confined structures with diameters below $\emptyset \leq 2.5$ µm. This is ascribed to an interplay between the stray field generated by the superconducting layer compared to the effective anisotropy field in confined geometries, which includes the dipolar field arising from the ferromagnetic Py regions situated between the structures. Indeed, as shown in Figure 1e,f, ($\square$, $\odot$)SC$^{\emptyset}$/FM$^{cont}$ structures exhibit a characteristic domain length comparable or larger than the lateral dimension of the largest imprinted Py region, which highlights the significant role played by the long-range dipolar field interactions originating from Py regions away from the SC structures. Structured Py samples have two benefits in this respect permitting a more efficient and a lower size-limit imprint. First, the SC stray field is maximized as they sit on top of the largest SC structures. Second, there are no FM regions in between Py structures and so no dipolar fields originating outside the SC/FM region that could compete with the SC magnetic stray field. The absence of an imprint for ($\square$, $\odot$)SC$^{20}$/FM$^{\emptyset}$ structures with $\emptyset < 2$ µm indicates however, an increase of the effective anisotropy field for smaller samples. We attribute this behavior to the presence of surface defects (CuO precipitates)[10]. Indeed, micromagnetic modelling show that the stray field originating from $\odot$SC$^{20}$ structures could allow the imprint of magnetic radial vortices in single crystal Py down to a lateral size of 900 nm (Supporting Information Section 3).

The presence of defects (Figure 1a-d) leads to local defective Py growth, resulting in non-magnetic regions within the FM layer, see Figure 1e-h. These defects can cause pinning and stabilization of the magnetic textures present within the spontaneous domain structure before the SC imprint. The pinning of the domain walls can lead to a reduction of the domain wall energy (local energy minimum) and to an increase of the coercivity[33] (Supporting Information Section S4). This is not surprising, since previously studies in submicrometer geometries confirmed that defects like polycrystallinity and geometrical confinement, enhance pinning and stabilization of complex topologically non-trivial textures like vortices and skyrmion tubes[34]. Consequently, the effective anisotropy field increases, and larger magnetic fields are necessary to erase the initial spontaneous domain structure. The relative importance of surface defects increases as the size of the Py dots is reduced, and so the anisotropy field eventually surpasses the available SC stray field preventing the imprint of smaller radial vortices.

Strategies to overcome the increase in coercive field due to the presence of defects would require the elimination of defects, when possible, or increasing the SC stray field. The later could be achieved either by increasing the size of the SC structures, and/or by increasing the critical current of the SC. Worth to mention that in this case the gain in size reduction would still be limited by the size and spatial distribution of defects as well as hindered by the tendency of the magnetic ground state to evolve towards a normal vortex state[35].

For those structures for which the SC stray field surpasses the anisotropy field, the presence of surface defects can exert a beneficial impact on the stabilization of the imprinted swirling domains across a wide temperature range. Increasing the temperature above $T_C$ leads to the disappearance of the SC stray field which stabilizes the imprint. For defect-free samples this leads to the magnetic relaxation of the system towards a normal vortex state (Supplementary information Section 3) which core polarization is determined by the z-component of the SC stray field. Conversely, the presence of surface defects can partially stabilize the imprint (Supplementary information Section 5).

Indeed, the temperature dependent XMCD data shown in Figure 4 reveal that despite some relaxation, individual structures retain above $T_C$ a memory of the radial vortex imprinted state. This is evidenced in panels a) and b) of Figure 5 where the local magnetization direction averaged for (□,⊙)SC$^{20}$/FM$^{20}$ structures at 140 K (T > $T_C$) after the low temperature SC imprint is shown by means of arrows. For both squares and discs, the angular distribution resembles that of a radial vortex despite no SC stray-field present.

Further details concerning the particularities of the SC imprint and memory effect for ▫ and ⊙ structures can be obtained by locally comparing the angular difference between the local magnetization direction experimentally measured at 140 K (Figures 5a,b) and that expected for a radial vortex magnetic distribution. This angular deviation from the ideal imprint ($\theta_{deviation}$), due to the suppression of the SC stray field, is represented in the 2D maps of Figure 5a,b by a color scale gradient.

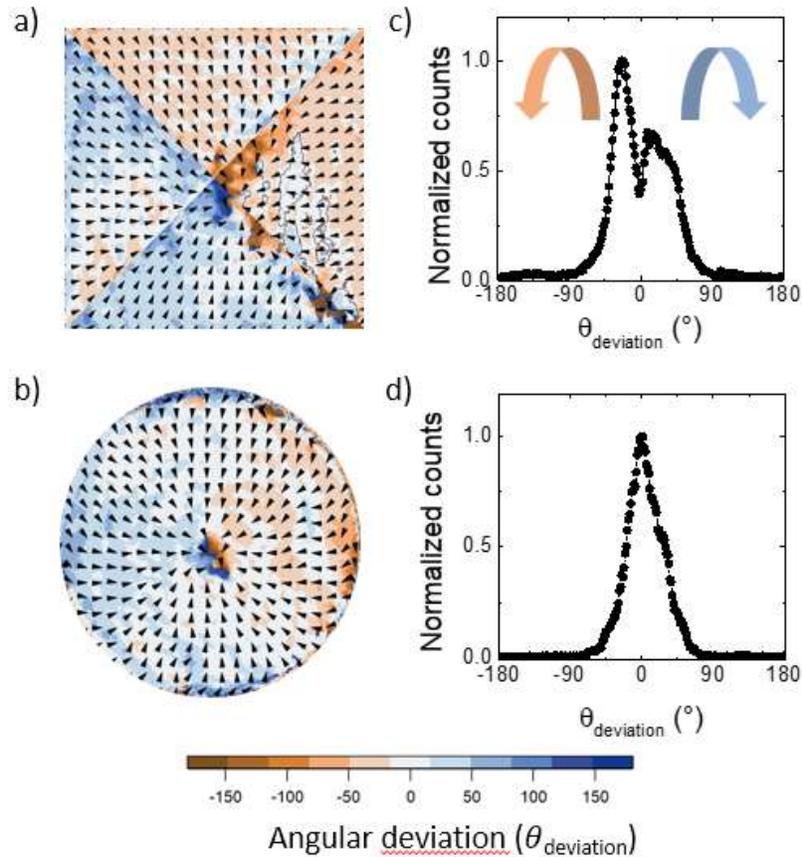

**Figure 5** a) and b) Arrows: Angular deviation of the magnetization at 140 K after the SC imprint of a magnetic radial vortex at 50 K averaged over 10 similar (▫, ⊙)$SC^{20}/FM^{20}$ structures (see methods) in order to highlight their common behavior. Color: Angular deviation between the angular orientation at 140 K and that expected for an ideal radial vortex distribution. c) and d) histograms of the angular deviation represented in panels a) and b). Color arrows in panels c) indicate the direction of the angular deviation.

Square-like structures (Figure 5a) exhibit an asymmetric angular distribution of $\theta_{deviation}$ with two peaks centered at ± 25° (Figure 5c). On the other hand, disc-shaped structures (Figure 5b) present a much sharper distribution of $\theta_{deviation}$, with a single peak centered at 0° (Figure 5d). These differences can be attributed to the presence of magnetic domain walls along the diagonals in the square structures after the low temperature imprint. These domain walls, separating head-to-head magnetic domains, increase the dipolar energy. At T > T$_C$, in the absence of the stabilizing SC stray field, the magnetostatic energy is minimized in the vicinity of those domain by relaxing the imprinted magnetic domain state. Regions away from the diagonals, tend to be more stable (Figure 5a). In comparison, the symmetry of the discs leads to a magnetic domain imprint with no DWs as the radial magnetization rotates continuously around the center with no preferred magnetization orientation. Micromagnetic modelling confirms that the absence of DWs within the discs leads to an easier impression and to a higher stability of the imprinted state as compared to similar size square-like structures (Figure 5b and Supporting Information Figure S7).

**Conclusion**

Radial vortex magnetic domain configurations have been crafted in Py by means of the stray field generated by SC structures down to 2 μm lateral size. The in-plane components of this field account not only for their imprint but also for their stability at T < T$_C$. We show experimentally that hybrid SC/FM structures retain memory of the imprinted magnetic domain state at T > T$_C$ despite the disappearance of the stabilizing SC stray field. Micromagnetic modelling indicates that instead, the stability above the superconducting transition temperature is provided by microstructural defects. IN the absence of defects, the system would relax to a conventional magnetic vortex state with its polarity determined by the z-component of the stray-field of the superconductor.

Overall, disc-shaped structures provide an enhanced preservation of the imprint than square geometries due to the due to confinement and circular symmetry, preventing the formation of head-to-head 90ª magnetic domains otherwise energy-penalizing. Future work will be directed to optimize the introduction of defects (size, shape, number…) within the FM structures to improve the stabilization of the magnetic structures imprinted by the SC stray fields. Hybrid SC/FM heterostructures open an appealing direction for SC-field design and manipulation of magnetic textures in soft magnetic material, with great potential to shape spintronic applications based on magnetic textures.

**Methods**

Sample fabrication: Electron beam lithography was performed in a Raith50 module mounted on a Zeiss EVO 50 scanning electron microscopy instrument to obtain a square and disc pattern with different sizes. The first step was performed in a YBCO single layer using negative resist to cover parts of the layer, which was later chemically etched. A second lithography step was performed to define square and disc holes on top of the YBCO square and holes using positive resist. Py was grown on top of the sample and then a lift-off was performed to eliminate Py outside of devices.

PEEM imaging: x-ray PEEM is a magnetic and element selective technique with a resolution of ca. 30 nm. Unlike many other techniques (e.g. magnetic force microscopy), x-ray PEEM delivers direct information about the magnetization, and the element selectivity guarantees that the recorded magnetic information comes only from the element under investigation. This is important, as other techniques would also prove the magnetic field generated by the superconductor. Magnetic sensitivity arises from the difference in absorption of circularly polarized radiation with left and right helicity from a magnetic element[36].

Experiments were done at the PEEM station at the UE49/PGMa beam line of the synchrotron radiation source BESSY II of the Helmholtz-Zentrum Berlin[37]. The angle of incidence of the incoming radiation with respect to the sample surface was of 16°, which ensured a sizable projection of the in-plane magnetization of the Py layer along the beam propagation direction, which gives rise to the XMCD signal.

Magnetic imaging was always performed in zero external field after a magnetic field pulse. The maximum pulse amplitude ± 100 mT with a pulse duration of 0.5–1 s and increasing/decreasing field rates of 10 mT/s. XMCD Images were collected at the Fe $L_3$-edge (707 eV) for incoming circularly polarized radiation with right (σ+) and left (σ−) helicity respectively. A total of 30 images, each with a 3 s integration time, were collected per helicity. Each image was normalized to a bright field image and drift corrected before their averaging. The XMCD images were obtained as (σ- − σ+)/(σ- + σ+) where σ+ and σ− were the averaged images for right and left circular polarized radiation, respectively.

2D maps of the magnetization direction were computed from two XMCD images obtained at 0° and 90° azimuthal rotation of the sample. XMCD images at 0° and 90° were averaged over ten similar structures.

Calculation of the stray field of the SC structures: The stray magnetic field of disc- and square-shaped superconductor structures was calculated using a 3D finite-element model based on the H-formulation of Maxwell's equation as in previous works[10,38,39] and introduced in the micromagnetic simulation as an external magnetic field. A critical current for YBCO of $J_c=10^{11}$ A/m$^2$ was used.

Micromagnetic simulations: Micromagnetic simulations by means of Mumax3 [40–42] have been performed to investigate the role of defects in the stabilization of magnetic radial vortices imprinted in Py as the temperature is raised above the superconducting transition temperature of YBCO. The ferromagnetic domain state below and above the superconducting transition temperature of YBCO was simulated by considering the presence or absence of a superconductor magnetic stray field, respectively. Ferromagnetic Py disc- and squared-shaped structures with dimension alike as those reported within the manuscript were simulated with the effective parameters of the Py layer in proximity with YBCO, calculated in previous works[43] ($M_{sat}$=0.86 MA/m; $A_{ex}$=13pJ/m, α=3.9·10$^{-3}$). The presence of defects was imitated by the random inclusion of 400 holes (420 nm diameter) over an area of 20 μm X 20 μm.

**Aknowledgments**

The research leading to this result has been supported by the project CALIPSOplus under the Grant Agreement 730872 from the EU Framework Programme for Research and Innovation HORIZON 2020. L.M. acknowledges the European Commission -Horizon Europe for funding under the project ProteNano-MAG (MSCA 101067742) F.G. acknowledges financial support from Ministerio de Economía y Competitividad through the MAT2017-87134-C2-1-R project. J.E.V acknowledges projects French ANR-22-CE30-0020-01 "SUPERFAST" , Flag ERA ERA-NET "To2Dox" and Cost Action "SUPERQUMAP".

# Supplementary Information

# Size-dependence and high temperature imprinted by superconductor stray fields.

D. Sanchez-Manzano[1,2], G. Orfila[2], A. Sander[1], L. Marcano[3,4], F. Gallego[2], M. A. Mawass[3], F. Grilli[5], A. Arora[3], A. Peralta[2], F.A. Cuellar[2], J.A. Fernandez-Roldan[6], N. Reyren[1], F. Kronast[3], C. Leon[2], A. Rivera-Calzada[2], J.E. Villegas[1], J. Santamaria[2], S. Valencia[3,*].

[1] Unité Mixte de Physique, CNRS, Thales, Université Paris-Saclay, Palaiseau, France.

[2] GFMC. Dept. Física de Materiales. Facultad de Física. Universidad Complutense. 28040 Madrid, Spain.

[3] Helmholtz-Zentrum Berlin, Albert-Einstein Str. 15, 12489 Berlin, Germany.

[4] Dept. of Physics, Faculty of Science, University of Oviedo 33007 Oviedo, Spain.

[5] Karlsruher Institut für Technologie, Institut für Technische Physik, 76344 Eggenstein-Leopoldshafen, Germany.

[6] Helmholtz-Zentrum Dresden-Rossendorf e.V., Institute of Ion Beam Physics and Materials Research, 01328 Dresden, Germany;

*Corresponding author: sergio.valencia@helmholtz-berlin.de

## 1.- Stray field generated by the superconductor structures: lateral size dependence

The x, y and z components of the magnetic field generated by 250 nm thick YBCO superconductor squares after a magnetic field pulse of +100 mT has been computed as function of their lateral size (Ø), see methods for details. The superconducting current present within the YBCO structure after the magnetic field pulse leads to a space-dependent magnetic stray field generated by the SC which strength and direction varies from point to point, see Figures S1 and S2 where the z and x components of this field are depicted for Ø **= 20, 10, 5, 2, and 1 µm.**

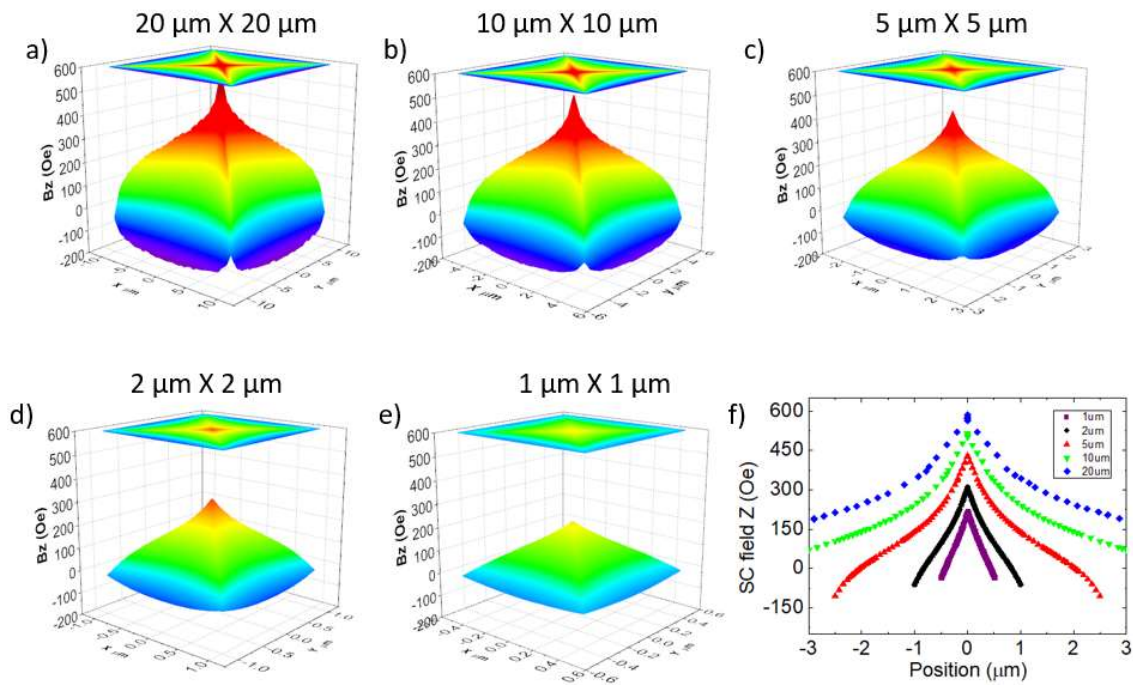

**Figure S1** a)-e) Z component of the strain field generated by a squared 250 nm thick SC structure after an out-of-plane magnetic field pulse of +100 mT. Lateral dimensions indicated on top of each panel. f) Profiles of the Z-component of the magnetic stray field of the SC measured from a straight line orthogonal to the edges and crossing the center. The reduction of the lateral size leads to an overall reduction of the magnetic stray field.

Devices for which the Py has been structured have been grown on top of YBCO structures with a lateral dimension of 20 µm. Figure S3 depicts x, y and z components of the magnetic stray field generated by a superconducting disc- and squared-shaped (⊙, ⊡)**SC$^\varnothing$.**

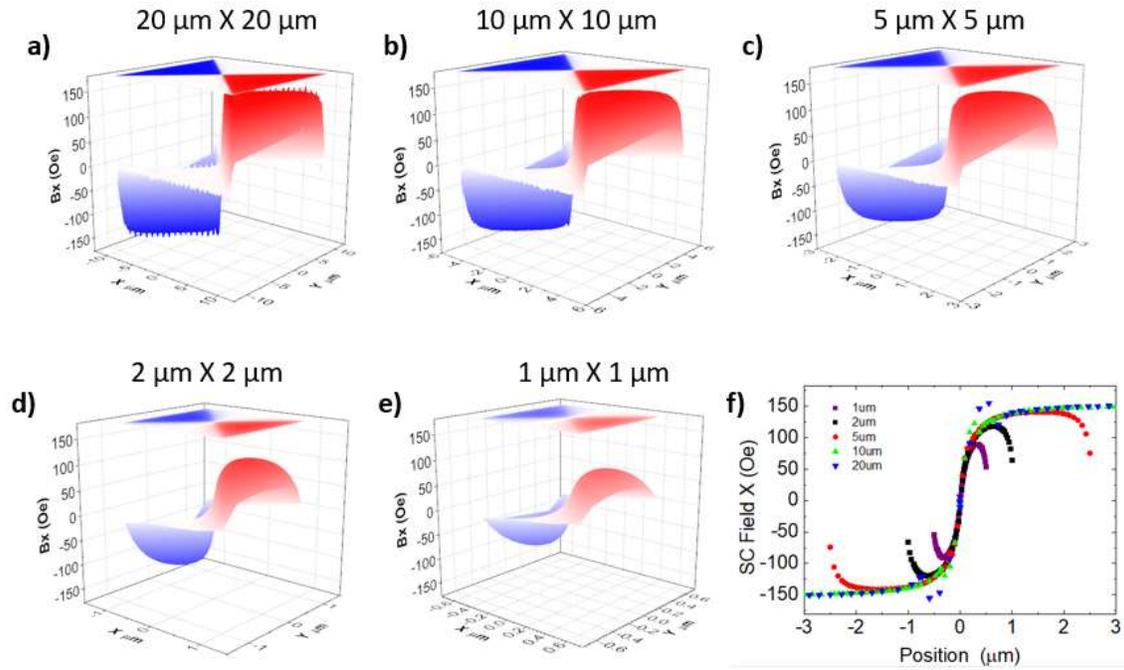

**Figure S2.** a)-e) X component of the SC stray-field for 250 nm thick YBCO square structures with lateral dimensions as indicated after an out-of-plane field pulse +100 mT. f) Profiles of the x-component of the magnetic stray field of the SC measured from a straight line orthogonal to the edges and crossing the center. The reduction of the lateral size leads to an overall reduction of the magnetic stray field. The y component shows the same behavior as the x one.

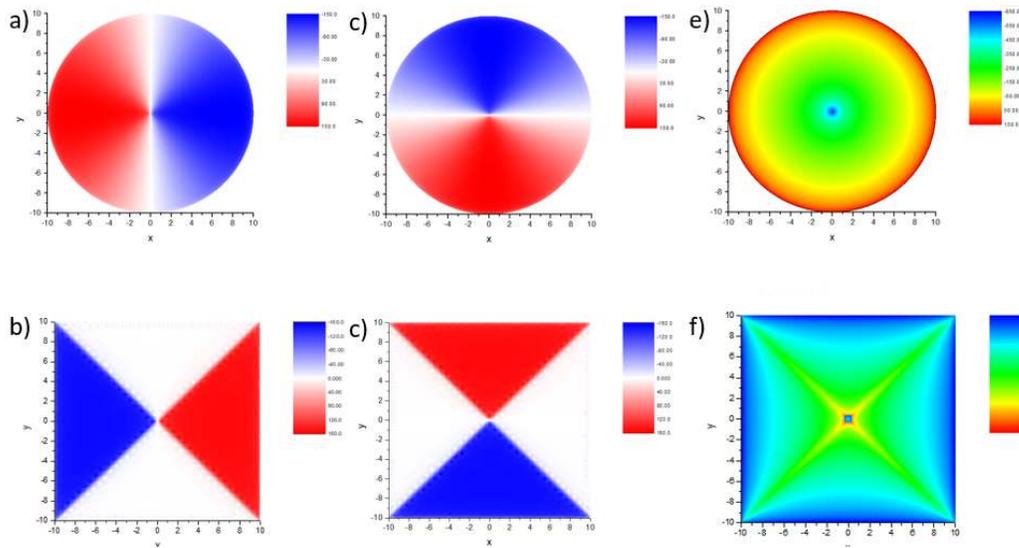

**Figure S3.** 2D representation of the X, y and Z components of the SC self-field (in Gauss) for a 20 µm size and 250 nm thick YBCO circular (a-c) and square (d-f) dot after an out-of-plane field pulse -100 mT (disc) and +100 mT (square).

## 2.- Differences in XMCD imaging of radial- and Landau- magnetic distribution.

In Landau vortex states, the magnetization curls (clockwise or anticlockwise) around the center of the structure to minimize the magnetostatic energy. In a radial vortex the magnetization direction points towards or away from the core on a radial direction orthogonal to the contour of the SC structure.

XMCD is proportional to the projection of the magnetization along the x-ray beam propagation direction. The XMCD signal is maximized for a magnetization direction parallel to the beam propagation direction.

Figure S4 sketches the expected XMCD images for radial and normal magnetic vortices measured with the incoming x-ray beam, along the blue arrow. Both, radial- and normal-vortex states show similar XMCD images but rotated by 90°.

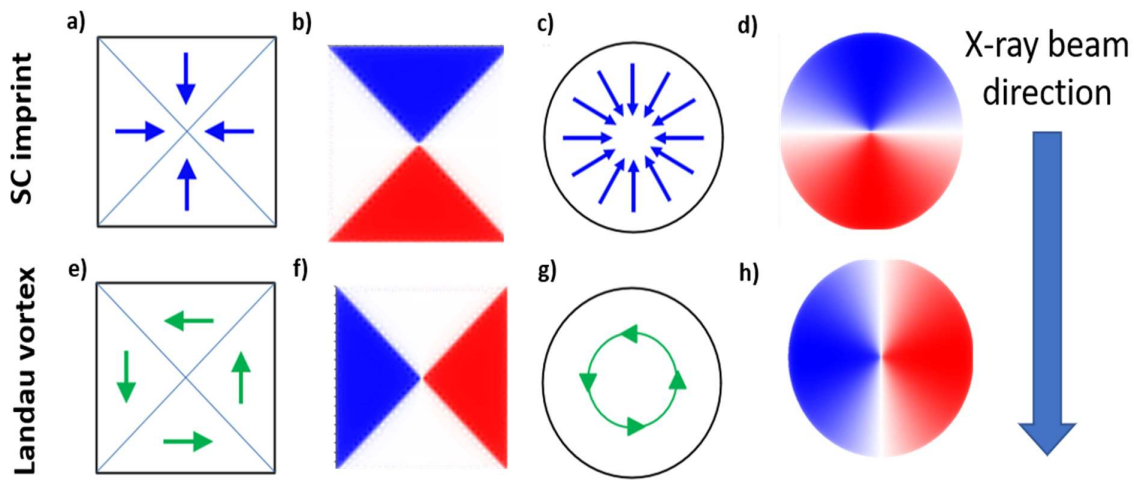

**Figure S4.** Comparison of the magnetization orientation distribution for radial vortex (panels a and c) and Landau or vortex-like magnetic domain distributions (panels e and h) for square and disc-shaped domains. Panels b and d and panels f and g simulated the expected XMCD images taking into account the experimental geometry, i.e. the orientation of the X-ray beam with respect to the structures.

## 3.- Relaxed vortex state for defect-free hybrid SC/FM disc-shaped structures after removal of the SC stray field.

We carried out finite difference micromagnetic modelling for hybrid $\odot SC^{20}/FM^{\emptyset}$ structures with no defects as function of the damping factor of Py to determine its magnetic domain state in the presence of the SC stray field and once this has been removed. Simulations have been performed

**for SC fields resulting from both +100 mT and -100 mT field pulses (see Figure S3) with a customized version of** Mumax 3[1–3] and postprocessing**.**

A prominent example of the switching process is depicted in figure S5, showing states before (Figure S5a), during (Figure S5b) and after (Figure S5c) the application of the SC field for a 3 µm diameter Py structure. Figure 5b evidences a radial distribution of the magnetization direction with a string distortion near the vortex core and at the dot boundaries. The distortion shapes the core in an spiral-like flavor in agreement to experimental XMCD images in Figure 4 Radial vortex-like states can be imprinted down to a lateral size of 900 nm (Figure S6).

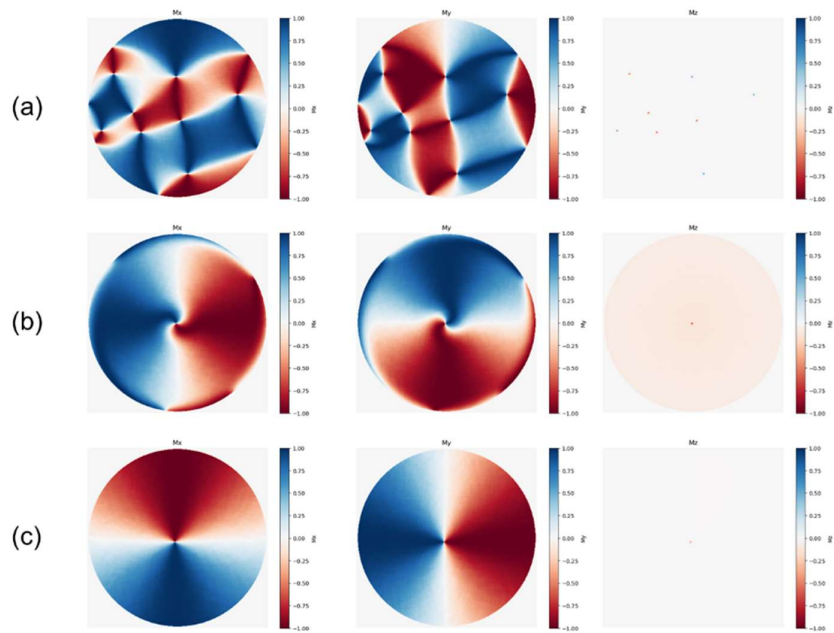

**Figure S5** Examples of selected magnetic states before the application of the SC field (a), during the SC field (b) and after removing the SC field (c) *in a Py disc with diameter of 3 µm. Each row shows the components x, y, z of the magnetization. The z- axis is perpendicular* to the disc plane.

**For all disc diameters investigated the suppression of the SC stray field leads to a normal vortex state where the magnetization curls around the center. Figure S7 summarizes the results from the simulations. The helicity of the resulting vortex state is independent of the damping factor and suggest that the helicity is likely to be determined by the microstructure. On the other hand, the polarity of the vortices is in most of the cases determined by the z-component (positive or negative) of the SC stray field (Figure S7a,b).**

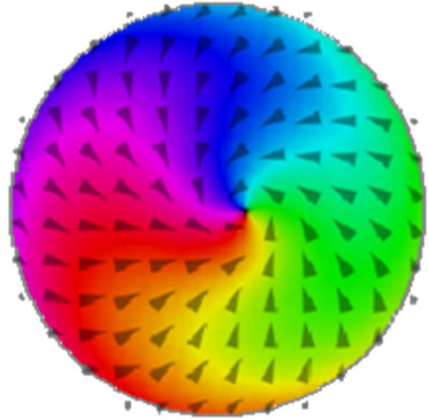

**Figure S6** 2D map of the magnetization orientation from micromagnetic simulations for defect-free Py ⊙ **structure of 900 nm diameter on top of a SC[20]** structure. Magnetic state in the presence of a SC resulting from a magnetic field pulse of +100 mT. The SC field allows the imprint of a radial vortex. Arrows indicate the local magnetization direction.

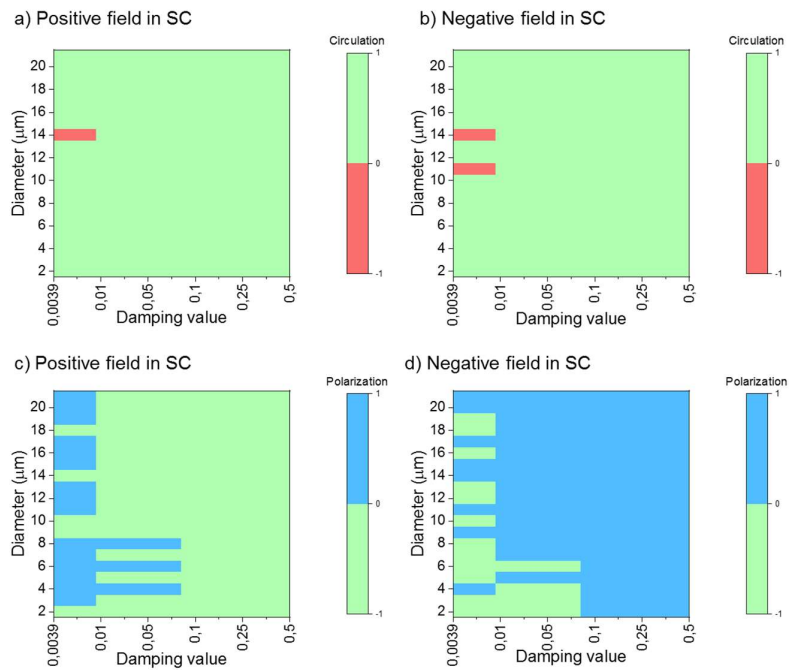

**Figure S7** Diagrams of the circulation and polarity of a vortex state in a dot with diameters between 1 and 20 µm after the application of either positive (a, c) or negative (b, d) currents in the superconductor. Values have been obtained solving Landau-Lifshitz-gilbert Equation with the SC field. The switching process is high dependent on the damping value close to realistic values of materials.

## 4.- Influence of surface defects for small structures.

To address the role played by surface defects we have simulated in-plane magnetic hysteresis loops, by means of Mumax3 micromagnetic simulations, for a Py disc-shaped structure 2.5 and 5 µm diameter with (main panel of Figure S8) and without (inset Figure S8) surface defects. While the extraction of quantitative information needs to be handled with care[1], a qualitative comparison between the results obtained for the different cases is feasible. Simulations corresponding to the defect-free sample show in both cases symmetric loops with relatively low coercive ($H_c$) and saturation fields ($H_s$). Coercive fields for structured Py are smaller than those produced by the YBCO superconducting structures of the same size and in particular smaller than that of the larger SC structure (Ø = 20 µm). The presence of defects leads to asymmetric loops with magnetization jumps associated to the nucleation and depinning of magnetic domains at defect positions. As a consequence, the coercive field increases by a factor 3 to 5 as compared to the defect-free case It is worth to remark that the size of the microstructures is large enough for the spontaneous formation of domain walls.

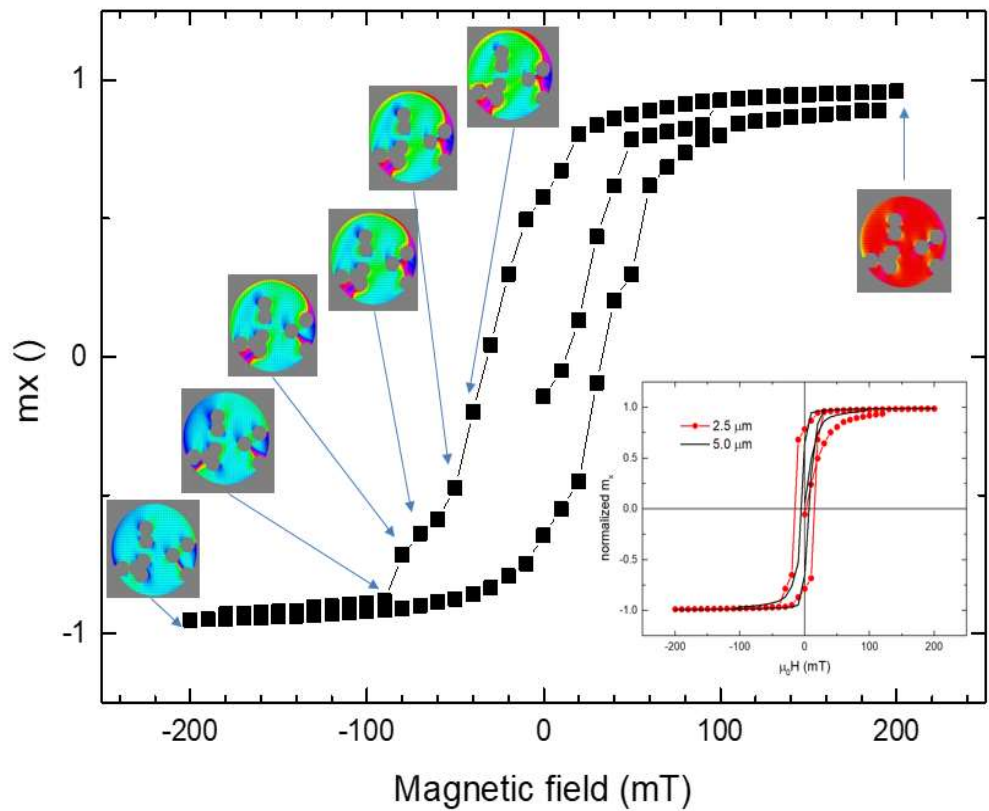

**Figure S8**. Main panel: Magnetic hysteresis loop (x component of the magnetization Vs in-plane field) simulated by means of Mumax 3 for a 6 nm thick circular $FM^{2,5}$ structure. The presence of defects leads to the pinning of magnetic domains which contribute to increase the coercive field as compared to structures with no defects (inset).

# 5.- Memory effect of imprinted magnetic radial vortices for hybrid SC/FM structures with defects.

Micromagnetic simulation by means of Mumax3[1–3] have been performed for a total of 10 (☐, ⊙)$SC^{20}/FM^{20}$ structures to determine the magnetic domain state after removal of the SC stray field. Each simulated structure contained 400 non-magnetic defects (diameter of 420 nm) aleatory distributed.

**Figure S9 depict 2D magnetization maps for single ☐ and ⊙ structures with the SC stray field "on". The magnetic domain state resembles the radial vortex magnetic state obtained for defect-free structures (Figure S5). Particularities of the magnetic state depend on the specific defect distribution. Panels c) and d) of Figure S9 shows the resulting magnetic domain state after the removal of the SC stray field. These images have been obtained by averaging all 10 similar structures, as done for the XMCD in the main manuscript, in order to determine their common behavior. Note that despite the disappearance of the SC stray field, which stabilizes the imprint, the resulting magnetic domain state resembles the radial vortex state imprinted by the superconductor. This is in contrast with what it is observed for defect-free structures for which the removal of the SC stray field leads to the system to relax to a vortex state.**

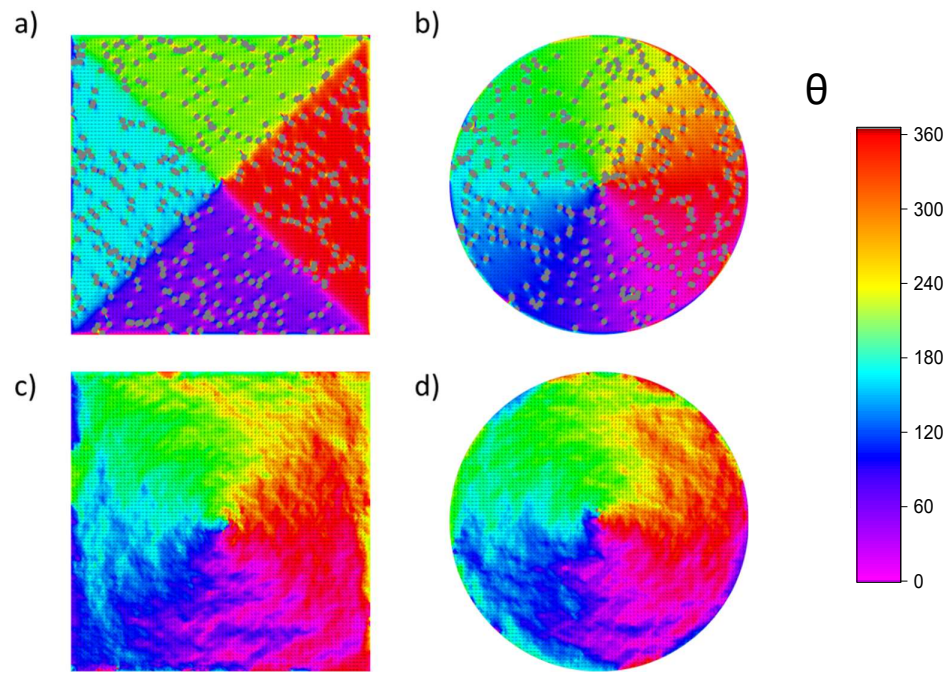

**Figure S9** 2D maps of the magnetization direction obtained from micromagnetic simulations for (☐, ⊙)$SC^{20}/FM^{20}$ structures with defects. a) and b) In the presence of the SC stray field the magnetic domain structure resembles that of a radial vortex c) and d) Magnetization maps resulting from the removal of the SC stray field averaged over 10 similar structures with aleatory defects distribution. The resulting magnetic domain distribution resembles that depicted in a) and b). That is, there is no relaxation to a vortex state as in the case of defect-free structures (**Figure S5**).

**Figure S10a,b shows 2D maps of the resulting magnetization direction (arrows) as well as of the deviation angle as compared to an ideal radial vortex. Panels c and d show the corresponding histograms. Results agree qualitatively with the experimental observations, thus highlighting the role of defects on the stabilization of the SC imprinted patter at T > Tc where the SC stray field is suppressed. Note larger deviation angles close to the diagonals of the square, positions where a domain wall is imprinted by the SC.**

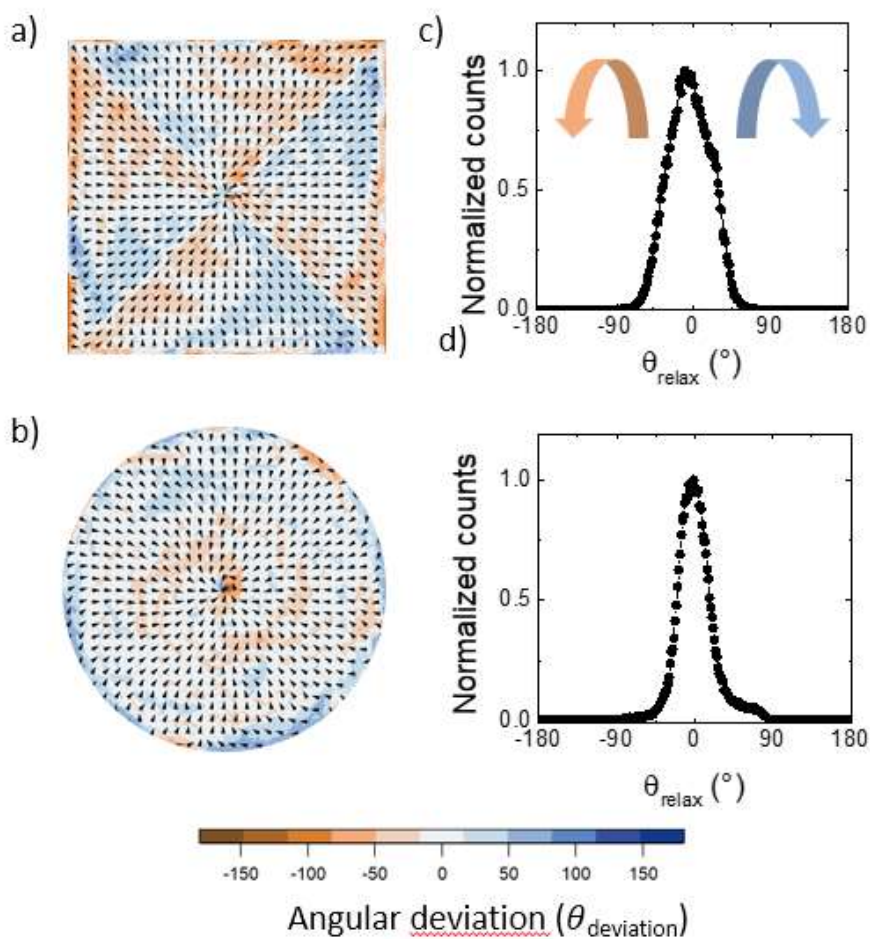

**Figure S10** Micromagnetic simulations: a) and b) Arrows: Angular orientation of the magnetization after suppression of the SC stray field after the SC imprint of a radial vortex magnetic domain state averaged over 10 similar (⊡,⊙)SC[20]/FM[20] structures (see methods) in order to highlight their common behavior. Color: Angular deviation between the angular orientation at 140 K and that expected for an ideal radial vortex. c) and d) histograms of the angular deviation represented in panels a) and b). Color arrows in panels c) indicate the direction of the angular deviation.